\documentclass[12pt]{article}
\usepackage{amssymb,amsmath,epsfig,graphicx}
\begin{document}
\title{\bf Existence of Static Wormhole Solutions in $f(R,G)$ Gravity}
\author{M. Farasat Shamir
\thanks{farasat.shamir@nu.edu.pk} and Saeeda Zia \thanks{saeeda.zia@nu.edu.pk} \\\\
Department of Sciences and Humanities, \\National University of
Computer and Emerging Sciences,\\ Lahore Campus, Pakistan.}
\date{}
\maketitle

\date{}

\maketitle
\begin{abstract}
This work investigates some feasible regions for the existence of traversable wormhole geometries in $f(R,G)$ gravity, where $R$ and $G$ represent the Ricci scalar and the Gauss-Bonnet invariant respectively. Three different matter contents anisotropic fluid, isotropic fluid and barotropic fluid have been considered for the analysis. Moreover, we split $f(R,G)$ gravity model into Strobinsky like $f(R)$ model and a power law $f(G)$ model to explore wormhole geometries. We select red-shift and shape functions which are suitable for the existence of wormhole solutions for the chosen $f(R,G)$ gravity model. It has been analyzed with the graphical evolution that the null energy and weak energy conditions for the effective energy-momentum tensor are usually violated for the ordinary matter content. However, some small feasible regions for the existence of wormhole solutions have been found where the energy conditions are not violated. The overall analysis confirms the existence of the wormhole geometries in $f(R,G)$ gravity under some reasonable circumstances.
\end{abstract}


\section{Introduction}

Current observations from different reliable resources and highly precise experimental data confirm that the universe is undergoing through an accelerated expansion (Spergel et al. 1997; Perlmutter et al. 2003). There have been different reasons in the literature suggesting this expansion ranging from the diverse models of dark energy to the alternative theories of gravity. Many valid questions regarding dark matter, initial singularity, cosmological constant and flatness issues may be effectively debated using modified theories of gravity. Therefore, the modified theories of gravity have provided the researchers with different cosmological approaches and thought to reveal the secrets behind the accelerated expansion of our universe. Taking into account the original theory, a number of alternative models have been proposed by constructing some complex lagrangians. For example, modified theories like $f(R)$, $f(R,T)$, $f(G$) and $f(R, G)$ have been structured with the combination of curvature scalars, topological invariants and their derivatives (B\"{o}hmer et al. 2007; Lattimer and Steiner 2014; Paliathanasis et al. 2014; Nojiri and Odintsov 2007; Capozziello et al. 2012; Astashenok et al. 2015; Capozziello et al. 2016; Felice et al. 2010; Bamba et al. 2012; Starobinsky 2007; Harko et al. 2011; Shamir and Zia 2017; Sharif and Zubair 2012; Shamir 2016). In $f(R,G)$ gravity, considering both $R$ and $G$ in terms of a non- linear combination provides an important extension of general relativity (GR). The quadratic behavior of $G$ in curvature invariants leads the early stages of evolution. Furthermore, with the introduction of $G$ along with $R$ provides the opportunity to get double inflationary situation (De Laurentis et al. 2015). Nojiri and Odintsov (2004) gave the idea of implicit and explicit couplings of the curvature term with the matter in $f(R)$ gravity. Sharif and Ikram (2016) presented a new modified $f(G,T)$ theory of gravity and studied energy conditions for Friedmann-Robertson-Walker universe. In the same theory, Shamir and Ahmad (2017) used the Noether symmetry approach with anisotropic background and investigated few cosmologically important $f(G,T)$ gravity models for locally rotationally symmetric Bianchi type-$I$ universe.

Wormhole can be thought of as imaginary topological features that give a channel for different space times away from each other. Wormholes are usually, classified as static and dynamic. In GR, an exotic fluid is required for creation of static wormholes which leads to violation of null energy condition (NEC). Investigating the wormhole solutions have been an interesting topic of discussion in the cosmological literature.
Morris and Thorne (1988) were the first to introduce the traversable wormhole. This idea was different from that previous concept of Einstein-Rosen bridge  (Einstein and Rosen 1935), and it was quite unlike to that of charge-carrying microscopic wormholes introduced by Wheeler (1962). For better understanding some interesting articles can be seen in(Hochberg and Visser 1998; Kim and Thorne 1991; Hawking 1992; Hochberg et al. 1997). In particular, the discussion of wormholes becomes more important in the context of alternative or modified theories of gravity. Lobo and Oliveira (2009) found the wormhole geometries in $f(R)$ gravity and gave some exact solutions by captivating specific shape-function and various equation of the state parameters. They gave the conclusion that violation of weak and null energy conditions are not necessary for formation of wormhole solutions in $f(R)$ gravity. Static wormhole solutions in $f(R)$ gravity were explored by Sharif and Zahra (2013) and they concluded that the wormhole solutions were possible in certain regions for only the barotropic matter case. Static spherically wormhole solutions have been discussed in $f(R,T)$ theory by Zubair et al. (2016). Sharif and Fatima (2015) investigated the wormholes in $f(G)$ gravity and provided some physically valid regions with isotropic and anisotropic matter distributions. Thus, it is would be an interesting to further investigate wormhole solution in modified gravity.

In this work, our intention is to explore some feasible regions for the existence of wormhole solution in $f(R,G)$ gravity. For this purpose, we split $f(R,G)$ gravity model into Strobinsky like $f(R)$ model and a power law $f(G)$ model. Furthermore, we have considered the isotropic, anisotropic and barotropic matter contents for a detailed analysis of the behavior of the shape function in connection with the weak and null energy conditions. The plan of our present work is as follows: in next section, we discuss the basic formalism of $f(R, G)$ gravity along with the corresponding structure of the field equations. In section $3$, we focus on the wormhole geometries in $f(R,G)$ gravity with three different types of matter contents. Section $4$ concludes and summarizes the findings of work.

\section{Some Basics of $f(R,G)$ Gravity}

An important modification of the theory has been proposed with the combination of scalar curvature with Gauss-Bonnet invariant known as $f(R,G)$ gravity (Cognola et al. 2006). It has been shown that $f(R,G)$ theory is reliable with the observational data and also stable. Furthermore, this theory explains well the accelerating waves of the celestial bodies, the phantom divide line crossing and evolution from acceleration to deceleration periods.
The action of modified Gauss-Bonnet gravity is defined as (Atazadeh and Darabi 2014)
\begin{eqnarray}\label{1}
S=\frac{1}{2K}\int d^4x\sqrt{-g}f(R,G)+S_M(g^{\mu\nu},\psi).
\end{eqnarray}
In this action, $f(R, G)$ is a bivariate function of $R$ and $G$.
Varying action (\ref{1}) with respect to metric tensor yields the following modified field equations
\begin{eqnarray}\label{2}
R_{\mu\nu}-\frac{1}{2}g_{\mu\nu}R&=&\kappa T^{(matt)}_{\mu\nu}+\nabla_\mu\nabla_\nu f_R-g_{\mu\nu}\Box f_R+2R\nabla_\mu\nabla_\nu f_G-2g_{\mu\nu}R\Box f_G\nonumber
\\&& -4R^\alpha_\mu\nabla_\alpha\nabla_\nu f_G
-4R^\alpha_\nu\nabla_\alpha\nabla_\mu f_G+4R_{\mu\nu}\Box f_G+4g_{\mu\nu}R^{\theta\phi}\nabla_\theta\nabla_\phi f_G\nonumber \\&&+4R_{\mu\theta\phi\nu}\nabla^\theta\nabla^\phi f_G-\frac{1}{2}g_{\mu\nu}V+(1-f_R)G_{\mu\nu},
\end{eqnarray}
where $f_R$ and $f_G$ are partial derivatives with respect to $R$ and $G$ respectively,
\begin{eqnarray}
V\equiv f_RR+f_GG-f(R,G),
\end{eqnarray}
and $T_{\mu\nu}^{(matt)}$ describes the ordinary matter.
The energy-momentum tensor in case of anisotropic fluid is given by
\begin{equation}\label{4}
T_{\alpha\beta}^{m}=(\rho+p_t)u_\alpha u_\beta-p_tg_{\alpha\beta}+(p_r-p_t)v_\alpha v_\beta,
\end{equation}
where $u_\alpha=e^{a/2} \delta_\alpha^0$, $v_\alpha=e^{b/2}\delta_\alpha^1$ are four velocities. Radial pressure and tangential pressures are $p_r$ and $p_t$ respectively, and the energy density is denoted by $\rho$.

\section{Wormhole Geometries and $f(R,G)$ Gravity}

A wormhole is a hypothetical tunnel with two ends that gives a subway across different space-times which are at long distances. The existence of a wormhole is due to the solution of Einstein field equations which provides a non-trivial structured linkage of dispersed points in the space-time.
The unification of the space and time into only a single space-time band, as proposed in the special theory of relativity, may let one theoretically to go across the space and time through a wormhole.

Here spherically symmetric wormholes will be discussed for three different matter distributions. For this purpose, we choose the spherically symmetric space-time as
\begin{eqnarray}\label{3}
ds^{2}=e^{a}dt^{2}-e^{b}dr^{2}-r^{2}{(d\theta^{2}+\sin^{2}\theta d\phi^{2})}.
\end{eqnarray}
where $a$ and $b$ are the arbitrary functions of radial coordinate $r$. Now we consider $b = log_e(1-\beta(r)/r)^{-1}$, where $\beta(r)$ is known as the shape function of the wormhole geometry. The term $a(r)$ in the metric coefficient represents the red-shift function which measures the magnitude of the gravitational red-shift. The geometry of the wormholes illustrates that the radial coordinate $r$ shows a non-monotonic behavior which gradually decreases from $  \infty \rightarrow r_0$ such that $\beta(r_0)=r_0=r$, with $r_0$ as the throat position, and then it increases from $r_0$ to $r\rightarrow \infty$. For the existence of traversable wormhole solutions the shape function $\beta(r)$ and the red-shift function $a(r$) must satisfy some important conditions. Firstly, the wormholes must not contain the event horizon, that is the red shift function $a(r)$ must not approach to $\infty$ at any point. Secondly, for a specific wormhole solution the flaring out condition of the wormhole throat, the inequality $\frac{(\beta-\beta')}{\beta^2}>0$ must be fulfilled at the throat $\beta(r_0)=r_0=r$. In addition to that, the condition $\beta'(r_0)<1$ also needs to be implemented to have the wormhole solutions. These are necessary requirements for the existence of the wormhole solutions. In GR, these conditions indicate the presence of exotic matter which requires the violation of the NEC. If the necessary conditions for the wormhole existence are not satisfied, situation may arise with the possibility of a black hole. In particular, solutions in GR allow the existence of wormhole where wormhole's throat is a black hole.
Now taking the equations (\ref{2}) and (\ref{4}) along with the metric (\ref{3}), the gravitational field equations in the form of energy density, the radial pressure and the tangential pressure are  as follows:
\begin{eqnarray}\label{5}
\rho&=&-e^{-b} f''_{1R}-\frac{4e^{-2b}}{r^2}(a'r-b'r-e^b+4)f''_{2G}-e^{-b}(\frac{b'}{2}+
\frac{2}{r})f'_{1R}+e^{-2b}(a''a'+\nonumber\\
&& a''b'-a''-a'^2+a'b'+\frac{a'^3}{2}-\frac{4a'}{r}-\frac{4b'e^b}{r^2}+\frac{8a''}{r}+\frac{6a'^2}{r}
-\frac{13b'}{r^2}-\frac{2b'^2}{r}-\nonumber\\
&&\frac{17a'}{r^2}-\frac{18e^b}{r^3}+\frac{18}{r^3})f'_{2G}+
\frac{e^{-b}}{r^2}(\frac{a''r^2}{2}+\frac{a'^2r^2}{2}-\frac{a'b'r^2}{4}+a'r)f_{1R}+\frac{e^{-2b}}{r^2}\nonumber\\
&&\{(1-e^b)(a'^2+2a''-a'b')-2a'b')\}f_{2G}-\frac{f}{2},
\end{eqnarray}
\begin{eqnarray}\label{6}
p_r&=&e^{-b}\big(\frac{a'}{2}+b'+\frac{2}{r}\big)f'_{1R}+e^{-2b}\{(2a''b'+a'^2b'-a'b'^2+
\frac{4a'b'}{r}+\frac{8a'}{r^2}-\frac{4b'^2}{r}-\nonumber\\
&&\frac{4a'e^b}{r^2}-\frac{4b'e^b}{r^2} -\frac{8e^b}{r^3}+\frac{2a'}{r^2}+\frac{4b'}{r^2}+\frac{8}{r^3})-r(2a''b'+a'^2b'-a'b'^2-
\frac{4b'^2}{r})\}f'_{2G}\nonumber\\&&-\frac{e^{-b}}{r^2}(\frac{a''r^2}{2}+\frac{a'^2r^2}{4}-
\frac{a'b'r^2}{4}-b'r)f_{1R}-\frac{e^{-2b}}{r^2}\{(1-e^b)(a'^2+2a''-a'b')-\nonumber\\&& 2a'b'\}f_{2G}+\frac{f}{2},
\end{eqnarray}
\begin{eqnarray}\label{7}
p_t&=&e^{-b} f''_{1R}+e^{-2b}(2a''+a'^2-a'b'+\frac{2a'}{r}-\frac{2b'}{r})f''_{2G}+e^{-b}(\frac{a'}{2}+
\frac{b'}{2}+\frac{1}{r})f'_{1R}\nonumber
\\&&-\frac{1}{2r^3}\{e^{-b}(2b'^2r^2+4b'r-4a'^2r^2-16a'r-24-2a'a''r^3+a'^2b'r^3-a'^3r^3)\nonumber
\\&&+32-8e^b\}f'_{2G}-\frac{e^{-b}}{r^2}(\frac{a'r}{2}-\frac{b'r}{2}-e^b+1)f_{1R}-\frac{e^{-2b}}{r^2}\{(1-e^b)(a'^2+2a''-a'b')\nonumber
\\&&-2a'b'\}f_{2G}+\frac{f}{2}.
\end{eqnarray}
Here prime denotes the derivatives with respect to the radial coordinate. The above expressions $\rho$, $ p_r$ and $p_t$ involving the matter threading wormholes consist of the shape function and the function $f(R,G)$ with its radial derivatives.
It can be observed that the equations (\ref{5})-(\ref{7}) are highly non linear and very much complicated. Therefore, it is difficult to find explicit forms of $\rho$, $p_r$ and $p_t$. For this particular situation, we split $f(R,G)$ gravity model into Strobinsky like $f(R)$ model and a power law $f(G)$ model
\begin{equation}\label{13}
f(R,G)=f_1(R)+f_2(G),
\end{equation}
where $f_1(R)$ is a function of Ricci scalar $R$  and $f_2(G)$ is a function the Gauss-Bonnet term $G$. Further, we assume $f_1(R)=R+\lambda R^2$, where $\lambda$ being an arbitrary constant and $f_2(G)=G^n$, $n\neq 0$.
Thus, using the $f(R,G)$ model (\ref{13}), the equations (\ref{5})-(\ref{7}) take the form
\begin{eqnarray}\label{5a}
\rho&=&-2\lambda e^{-b} R''-\frac{4n(n-1)(n-2)e^{-2b}}{r^2}(a'r-b'r-e^b+4)G^{n-3}G''-2\lambda e^{-b}\nonumber\\
&&(\frac{b'}{2}+\frac{2}{r})R'+n(n-1)e^{-2b}(a''a'+a''b'-a''-a'^2+a'b'+\frac{a'^3}{2}-\frac{4a'}{r}-\nonumber\\
&&\frac{4b'e^b}{r^2}+\frac{8a''}{r}+\frac{6a'^2}{r}-\frac{13b'}{r^2}-\frac{2b'^2}{r}-\frac{17a'}{r^2}-\frac{18e^b}{r^3}+\frac{18}{r^3})G^{n-2}G'+\nonumber\\
&&\frac{e^{-b}}{r^2}(\frac{a''r^2}{2}+\frac{a'^2r^2}{2}-\frac{a'b'r^2}{4}+a'r)(1+2\lambda R)+\frac{ne^{-2b}}{r^2}\nonumber\\
&&\{(1-e^b)(a'^2+2a''-a'b')-2a'b'\}G^{n-1}-\frac{1}{2}(R+\lambda R^2+G^n),
\end{eqnarray}
\begin{eqnarray}\label{6a}
p_r&=&2\lambda e^{-b}\big(\frac{a'}{2}+b'+\frac{2}{r}\big)R'+n(n-1)e^{-2b}\{(2a''b'+a'^2b'-a'b'^2+\frac{4a'b'}{r}+\nonumber\\
&&\frac{8a'}{r^2}-\frac{4b'^2}{r}-\frac{4a'e^b}{r^2}-\frac{4b'e^b}{r^2}-\frac{8e^b}{r^3}+\frac{2a'}{r^2}+\frac{4b'}{r^2}+\frac{8}{r^3})-r(2a''b'+a'^2b'-\nonumber\\
&&a'b'^2-\frac{4b'^2}{r})\}G^{n-2}G'-\frac{2\lambda e^{-b}}{r^2}(\frac{a''r^2}{2}+\frac{a'^2r^2}{4}-\frac{a'b'r^2}{4}-b'r)(1+2\lambda R)-\nonumber\\
&&\frac{ne^{-2b}}{r^2}\{(1-e^b)(a'^2+2a''-a'b')- 2a'b'\}G^{n-1}+\frac{1}{2}(R+\lambda R^2+G^n),
\end{eqnarray}
\begin{eqnarray}\label{7a}
p_t&=&2\lambda e^{-b} R''+n(n-1)(n-2)e^{-2b}(2a''+a'^2-a'b'+\frac{2a'}{r}-\frac{2b'}{r})G^{n-3}G''+\nonumber
\\&&2\lambda e^{-b}(\frac{a'}{2}+\frac{b'}{2}+\frac{1}{r})R'-\frac{n(n-1)}{2r^3}\{e^{-b}(2b'^2r^2+4b'r-4a'^2r^2-16a'r-24-\nonumber
\\&&2a'a''r^3+a'^2b'r^3-a'^3r^3)+32-8e^b\}G^{n-2}G'-\frac{e^{-b}}{r^2}(\frac{a'r}{2}-\frac{b'r}{2}-e^b+1)\nonumber
\\&&(1+2\lambda R)-\frac{2e^{-2b}}{r^2}\{(1-e^b)(a'^2+2a''-a'b')-2a'b'\}G^{n-1}+\frac{1}{2}(R+\nonumber
\\&&\lambda R^2+G^n).
\end{eqnarray}

The violation of the energy conditions in the presence of exotic matter has been the essential and fundamental reason behind the existence of the wormhole geometries in the classical GR. However, the modified theories of gravity have shown some different characteristics to some extent. It is believed that these energy conditions may be proposed due to the existence of a term $R_{\alpha\beta}\mu^\alpha\mu^\beta$ in the Raychaudhuri equations (Raychaudhuri 1957)for the time-like congruences and null geodesics denoted by $\nu^\alpha$  and  $\mu^\alpha$ respectively. The equations are
\begin{equation}
\frac{d\theta}{d\tau}=\omega_{\alpha\beta}\omega^{\alpha\beta}-\sigma_{\alpha\beta}\sigma^{\alpha\beta}-R_{\alpha\beta}\nu^\alpha\nu^\beta-\frac{\theta^2}{3},
\end{equation}
\begin{equation}
\frac{d\theta}{d\tau}=\omega_{\alpha\beta}\omega^{\alpha\beta}-\sigma_{\alpha\beta}\sigma^{\alpha\beta}-R_{\alpha\beta}\mu^\alpha\mu^\beta-\frac{\theta^2}{2},
\end{equation}
 where the terms $\theta$, $\omega_{\alpha\beta}$  and $\sigma^{\alpha\beta}$ stand for the expansion scalar, rotation tensor and the shear tensor respectively. These terms are connected to the congruences of the null and time-like geodesics.
The NEC $T^{}_{\alpha\beta}\mu^\alpha\mu^\beta\geq0$ are obtained after manipulating the field equations (\ref{2}) using Raychaudhuri equations.
The energy conditions are defined as
\begin{eqnarray}\nonumber
\\&&\mathrm{NEC}:~~~~~~\rho^{}+p^{}_r\geq0,~~~\rho+p^{}_t\geq0\nonumber,
\\&&\mathrm{WEC}:~~~~~~\rho^{}\geq0,~~~\rho^{}+p^{}_r\geq0,~~~\rho+p^{}_t\geq0\nonumber,
\\&&\mathrm{SEC}:~~~~~~\rho^{}+p^{}_r\geq0,~~~\rho^{}+p^{}_t\geq0,~~~\rho^{}+p^{}_r+2p^{}_t\geq0\nonumber,
\\&&\mathrm{DEC}:~~~~~~\rho^{}>\mid p^{}_r\mid,~~~\rho>\mid p^{}_t\mid.\nonumber
\end{eqnarray}
where NEC, WEC, SEC and DEC are the null energy conditions, weak energy conditions, strong energy conditions and dominant energy conditions respectively.
Violation of NEC in GR is the fundamental property for traversable wormhole because its violation leads to the violation of all energy conditions.
For the study of wormhole solutions in different modified theories of gravity, the redshift function have been chosen to be constant in many cases which proves to be quite helpful in simplifying the field equations. But in our case while studying the model (\ref{13}) with constant redshift function, our gauss-bonnet term vanishes as it was in the case of the $f(G)$ gravity (Sharif and Ikram 2015). Therefore, to deal with this issue, we choose a non-constant redshift function of the following form (Kar and Sahdev 1995; Anchordoqui et al. 1998),
\begin{equation}\label{14}
a=-\frac{2\eta}{r},~~~~~~~~~~~~~~\eta > 0.
 \end{equation}
It is to be noted here that the term $a$ remains finite and non-zero for all $r> 0$. This suggests that there would be no horizons. Moreover, it can be seen from its expression that $a\rightarrow0$ as $r\rightarrow\infty $ which shows that $a$ has to be asymptotically flat as well.
NEC violations by the effective energy momentum tensor throughout the wormhole throats occur due to the satisfaction of eq. (\ref{14}).

\subsection{Anisotropic Fluid}

For anisotropic fluid, we choose the following shape function (Boehmer et al. 2012; Jamil et al. 2013; Bhattacharya and Chakraborty 2017)
\begin{equation}\label{8}
b(r)=-ln[1-(\frac{r_0}{r})^{m+1}],
\end{equation}
which gives

\begin{equation}\label{9}
\beta(r)=(r_0)^{1+m}r^{-m},
\end{equation}
where $r_0$ and $m$ are the real arbitrary constants.
Clearly, $\beta(r)$ depends on the value of $m$. It can be seen that the different choices for the values of $m$ will result into the different forms of the shape function. For the existence of the shape function $\beta(r)$ must meet the flaring out condition i.e.  $\beta'(r) <1$. This also trivially satisfies the condition of $\beta(r_0)= r_0$.
For example, if we choose $ m =\frac{-1}{2},\frac{1}{2}, 1, -3$ etc, we have $\beta(r)$ as $\sqrt{r_0r}$, $r_0^{\frac{3}{2}} r^{\frac{-1}{2}}$, $\frac{r^2_0}{r}$ and $r^2_0r^3$ respectively. Using the same shape function (\ref{9}) for $m = \frac{-1}{2}$ and $m = 1$, Lobo and Oliveria (2009) investigated the wormhole solutions in $f(R)$ gravity. Pavlovic and Sossich (2015) discussed the wormhole geometries without exotic matter for the different models in $f(R)$ gravity by taking $m =\frac{1}{2}$  . In a recent paper, Zubair et al. (2016) investigated the static spherically symmetric wormhole solutions in $f(R,T)$ gravity and discussed in detail the feasible regions for the energy constraints by implementing the same shape function for $m =\frac{1}{2}$. Furthermore, the condition for the asymptotically flat metric that is $r^{-1}\beta\rightarrow 0$ as $r$ approaches to $\infty$, is also satisfied by the shape function (\ref{9}).
Taking into account the expressions for the red shift function $a(r)$ and the shape function $b(r)$, and after working with equations (\ref{5})-(\ref{7}), we obtain the important expressions for $\rho$, $ p_r$ and $p_t$. As these new expressions are very long, so related graphs are shown in Figs.(1) and (2).
\begin{figure}\center
\begin{tabular}{cccc}
\epsfig{file=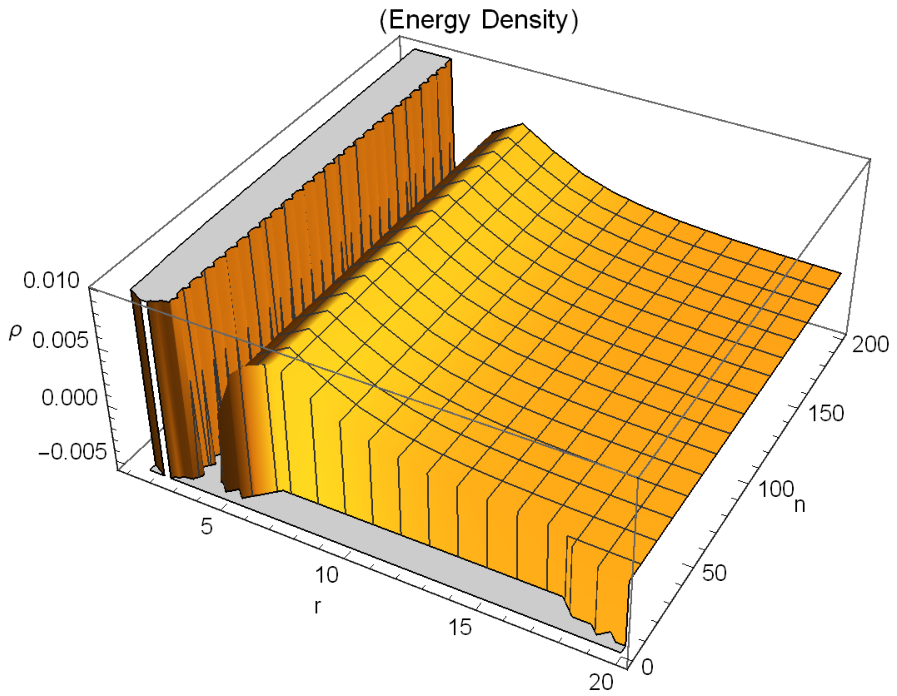,width=0.35\linewidth} &
\epsfig{file=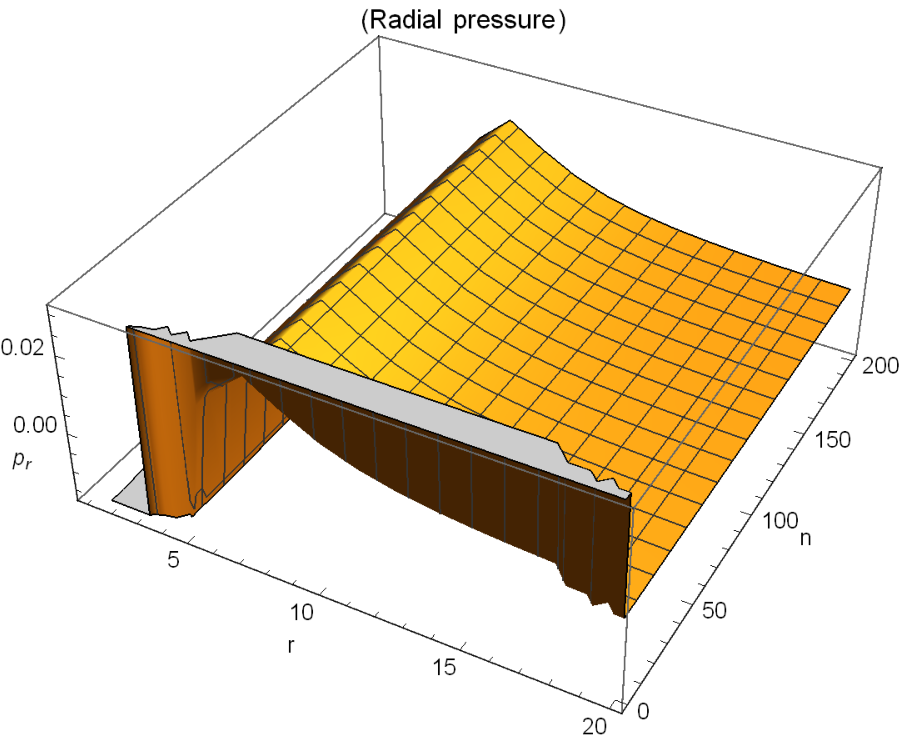,width=0.35\linewidth} &
\epsfig{file=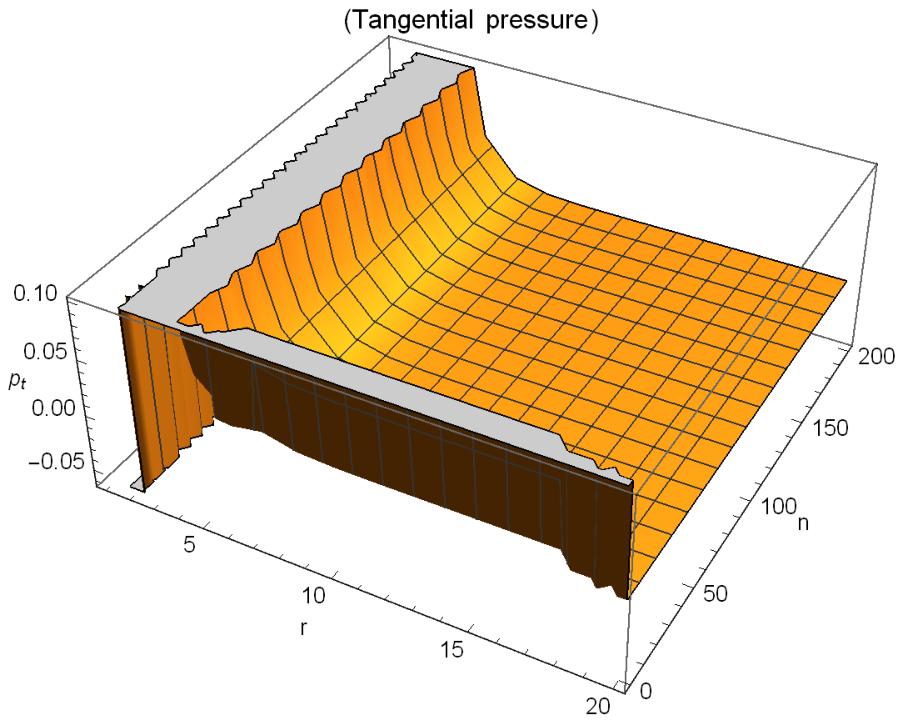,width=0.35\linewidth} \\
\end{tabular}
\caption{Variation of $\rho~~(MeV/fm^3)$, $p_r~~(MeV/fm^3)$ and $p_t~~(MeV/fm^3)$ with radial coordinate $r (Km)$ and model parameter $n$, $m=\frac{1}{2}$, $\eta=5$}\center
\end{figure}
Here we have taken $r_0 = 1~~(MeV/fm^3)$. For the parameters $\eta$ and $m$, we need to choose their suitable values for the validity of WEC and NEC to find the desired regions for the existence of the wormhole geometries in $f(R,G)$ model. We have investigated the possibility of feasible regions for the wormhole geometries by taking $m = -3, \frac{-1}{2}, \frac{-1}{2}, 1$. However, our discussions remain focused mainly for $m =\frac{1}{2}, -3$  and $\eta = 5$ to show up the illustrations. We are putting here a detailed analysis with the help of $3D$ valid regions of the corresponding energy conditions.
\begin{figure}\center
\begin{tabular}{cccc}
\epsfig{file=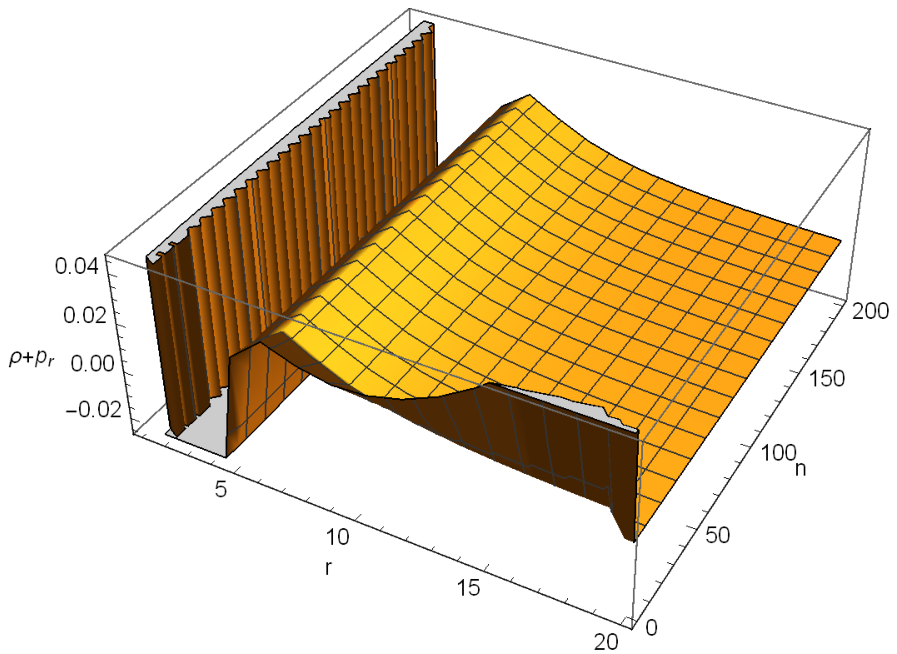,width=0.35\linewidth} &
\epsfig{file=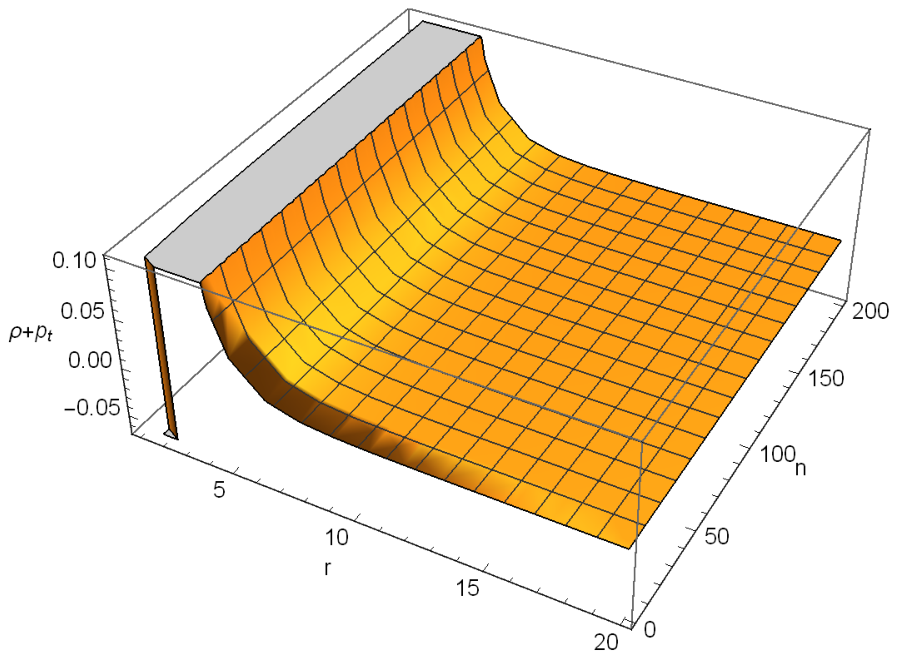,width=0.35\linewidth} \\
\end{tabular}
\caption{Behaviour of $\rho+p_r~~(MeV/fm^3)$ and $\rho+p_t~~(MeV/fm^3)$ with radial coordinate $r (Km)$ and model parameter $n$, $m=\frac{1}{2}$, $\eta=5$}\center
\end{figure}
We find different situations for WEC and NEC validity regions with the following restrictions.\\\\
$\bullet$ Let us first consider the case when $m =\frac{1}{2}$, $n=2$ and $\eta=5$.
When $4.5\leq r\leq10$, $\rho,~\rho+p_t,~\rho+p_r$ all quantities are greater than zero, both NEC and WEC are valid and these conditions violate when $r\geq10$. For small regions like $0<r<0.002$, $0.9<r\leq1$, $1.3<r<2.2$ and for a large region $r>10$, NEC and WEC are violated because $\rho+p_t<0$. For region when $2.3<r<4.5$, $\rho+p_r \geq0$ is not fullfilled so violation of NEC and WEC can be declared.\\\\
$\bullet$ Next, we consider the case when $m =-3$, $n=2$  and $\eta=5$, which implies that $\beta=r_0^2r^3$.
When $1\leq r \leq 1.25$, all three quantities $\rho,~\rho+p_t,~\rho+p_r$ are positive, both NEC and WEC are simultaneously valid. For small region like $1.5<r<2$, $\rho <0$ but $\rho+p_t>0$ and $\rho+p_r>0$, so only NEC is valid here. For region when $r>10$ only $\rho+p_r>0$, this confirms the violation of both NEC and WEC.\\\\
$\bullet$ It is worthy to mention here that results for other choices for different values $m=\frac{-1}{2},~\frac{1}{5},~1$ also give results similar to the case discussed in detail for $m=\frac{1}{2}$. The only difference is in the bounds of $r$. All these choices either show validity or violation of energy conditions in different regions.

\subsection{Isotropic Fluid}

We consider here the specific case of isotropic condition $p = p_r = p_t$. Using equations (\ref{6a}) and (\ref{7a}) along with the isotropic condition, we obtain the following equation
\begin{eqnarray}\nonumber
\\&&-2\lambda e^{-b} R''-n(n-1)(n-2)e^{-2b}(2a''+a'^2-a'b'+\frac{2a'}{r}-\frac{2b'}{r})G^{n-3}G''+\nonumber
\\&&2\lambda e^{-b}(\frac{b'}{2}+\frac{1}{r})R'+n(n-1)\{e^{-2b}\big((2a''b'+a'^2b'-a'b'^2+\frac{4a'b'}{r}+\frac{8a'}{r^2}-\frac{4b'^2}{r}-\nonumber
\\&&\frac{4a'e^b}{r^2}-\frac{4b'e^b}{r^2}-\frac{8e^b}{r^3}+\frac{2a'}{r^2}+\frac{4b'}{r^2}+\frac{8}{r^3})-r(2a''b'+a'^2b'-a'b'^2-\frac{4b'^2}{r})\big)+\nonumber
\\&&\frac{1}{2r^3}\big(e^{-b}(2b'^2r^2+4b'r-4a'^2r^2-16a'r-24-2a'a''r^3+a'^2b'r^3-a'^3r^3)+\nonumber
\\&&32-8e^b\big)\}G^{n-2}G'-\frac{e^{-b}}{r^2}(\frac{a''r^2}{2}+\frac{a'^2r^2}{4}-\frac{a'b'r^2}{4}-b'r+\frac{a'r}{2}-\frac{b'r}{2}-e^b+1)\nonumber
\\&&(1+2\lambda R)-2\frac{ne^{-2b}}{r^2}\{(1-e^b)(a'^2+2a''-a'b')- 2a'b'\}G^{n-1}=0.\label{15}
\end{eqnarray}
The equation (\ref{15}) is a non-linear homogenous differential equation. It is a very complicated equation for which an explicit expression of $\beta(r)$, through an analytic solution seems difficult. Therefore, we are left with the only possibility of using some numerical method, which gives us some results about the shape function, shown in Fig. $3$. The increasing behavior of the shape function is apparent from the evolution of the shape function. The condition $\beta-r< 0$ is fulfilled and $\frac{\beta}{r}$ goes to zero as $r\rightarrow\infty$, showing that the metric is asymptotically flat, a core condition for the existence of wormhole solutions. The throat is formed at $r_0=0.0101208$ satisfying the conditions $\beta(r_0=r_0)$ and $\beta'(r_0)<1$. The evolution of WEC and NEC has been shown in Fig. $4$. It can be noticed that both conditions are violated when $0<r\leq0.25$ but these conditions are satisfied afterwards. So wormholes can be formed in some specific regions in this case.
\begin{figure}\center
\begin{tabular}{cccc}
\epsfig{file=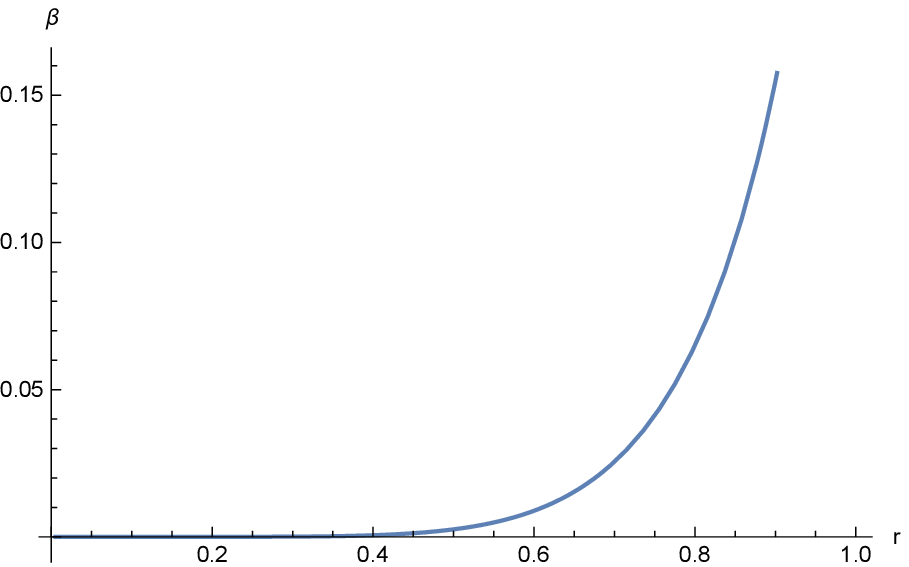,width=0.35\linewidth} &
\epsfig{file=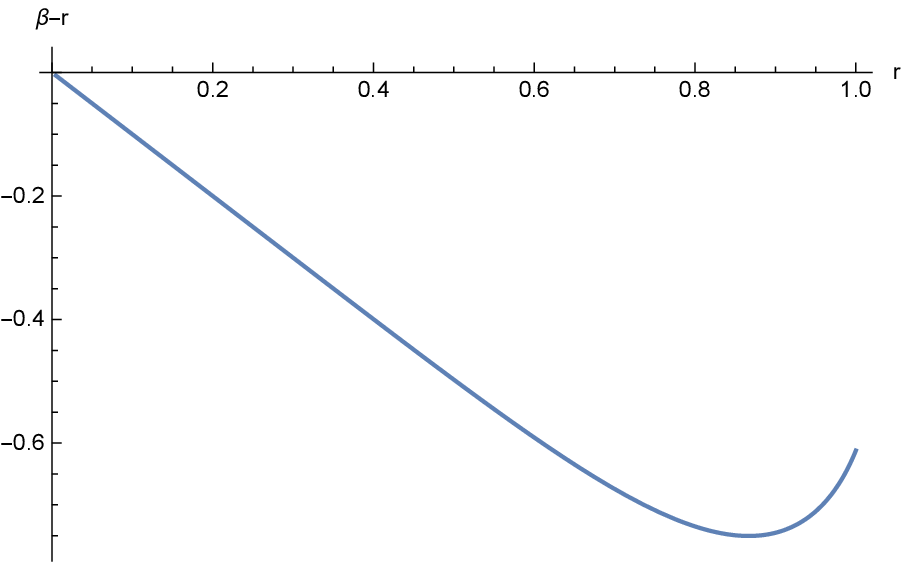,width=0.35\linewidth} &\\
\epsfig{file=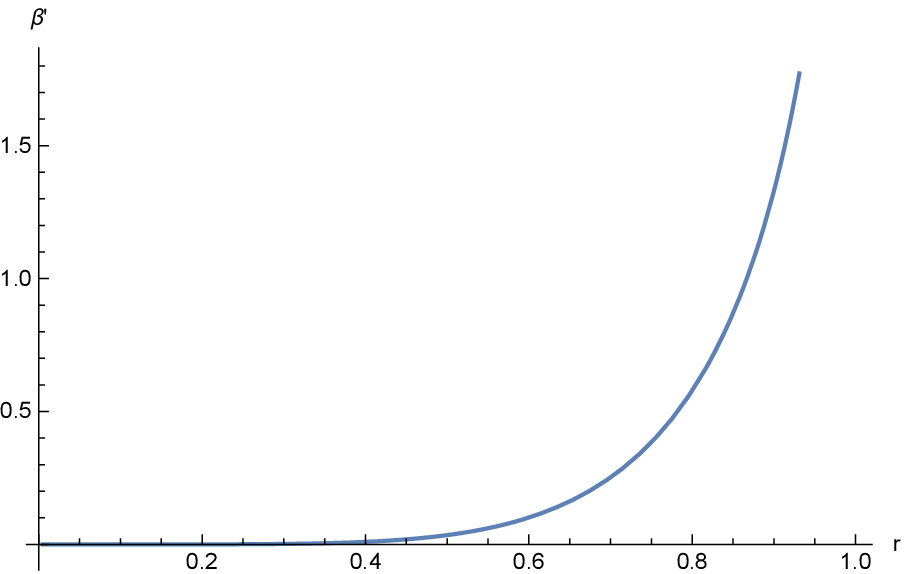,width=0.35\linewidth} &
\epsfig{file=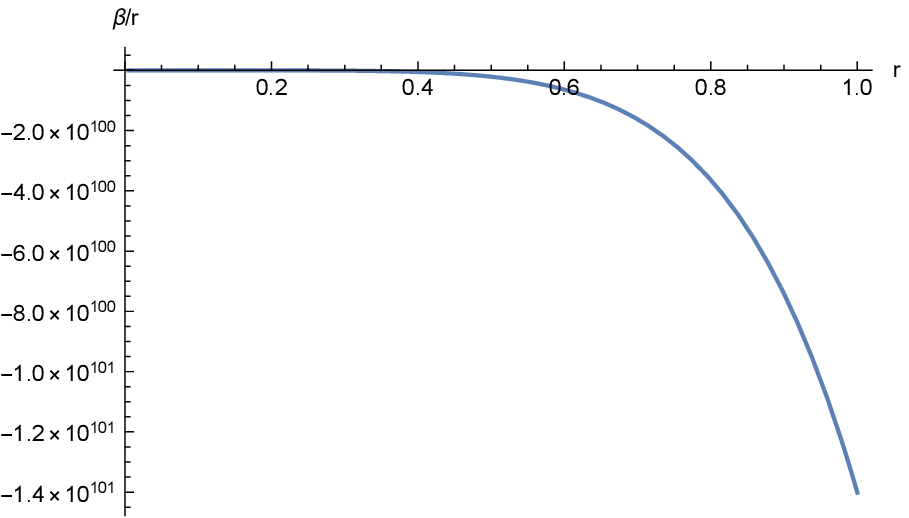,width=0.35\linewidth} \\
\end{tabular}
\caption{For isotropic fluid, behavior of $\beta(r)$, $\beta(r)-r$, $\beta'(r)$ and $\beta(r)/r$  with radial coordinate $r$. Here, we set  $n=2$, $\eta=5$ and $\lambda=1$}\center
\end{figure}
\begin{figure}\center
\begin{tabular}{cccc}
\epsfig{file=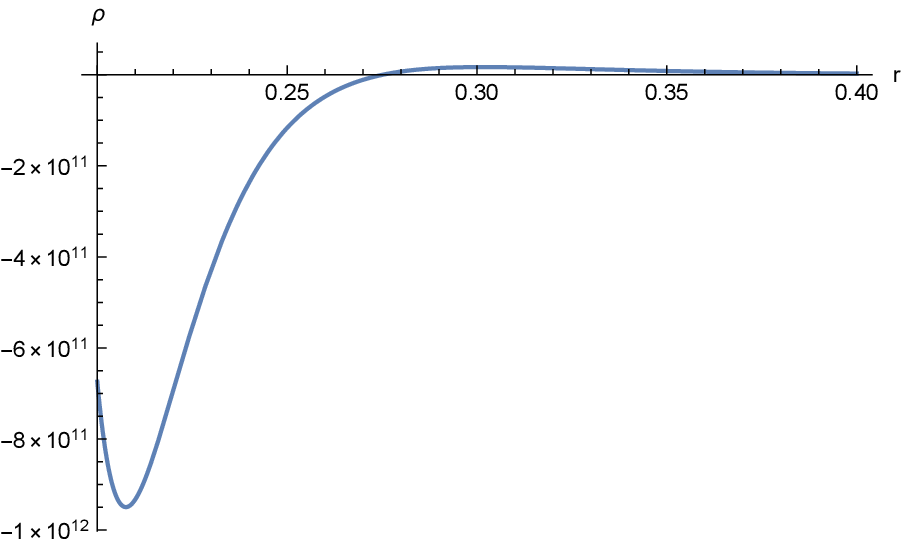,width=0.35\linewidth} &
\epsfig{file=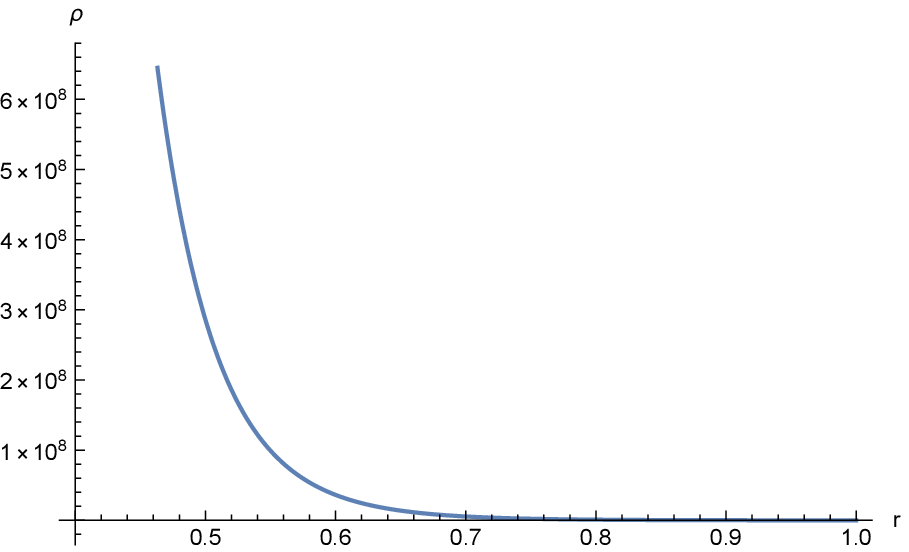,width=0.35\linewidth} &\\
\epsfig{file=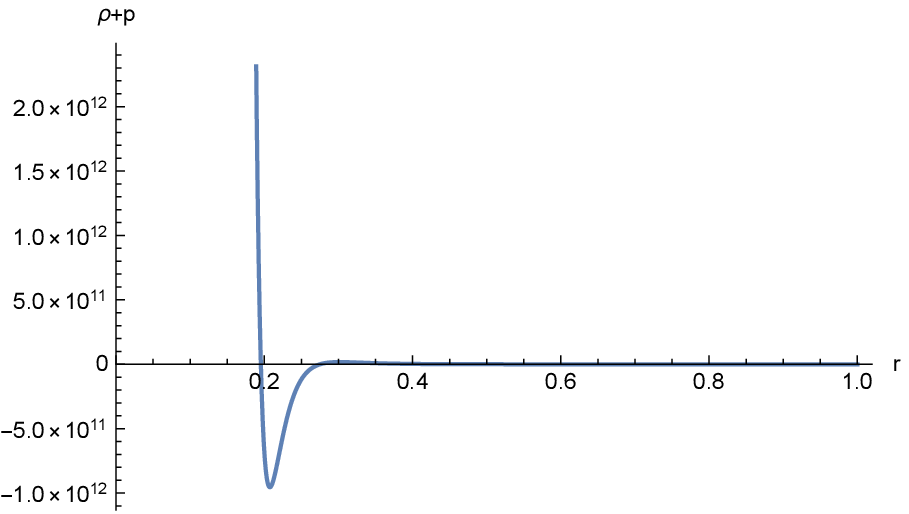,width=0.35\linewidth}& \\

\end{tabular}
\caption{For isotropic fluid, behavior of $\rho~~(MeV/fm^3)$ with radial coordinate $r~~(Km)$. Here, we set  $n=2$, $\eta=5$ and $\lambda=1$}\center
\end{figure}

\subsection{Barotropic Fluid}

We consider here testing of the radial pressure and the energy density relation with equation of state (EoS) parameter $\omega$, i. e., $p_r=\omega \rho$. Similar investigations through the EoS parameters have been done for $f(R, T)$, $f(G)$, $f(R)$, and $f(T)$ theories of gravity to study the wormhole geometries. Now in this case using corresponding expressions for $\rho$ and $p_r$ from equations (\ref{5a}) and (\ref{6a}), we obtain
\begin{eqnarray}\label{10}
&&2\lambda\omega e^{-b} R''+\frac{4\omega n(n-1)(n-2)e^{-2b}}{r^2}(a'r-b'r-e^b+4)G^{n-3}G''+\nonumber\\
&&2\lambda e^{-b}[\big(\frac{a'}{2}+b'+\frac{2}{r}\big)+\omega\big(\frac{b'}{2}+\frac{2}{r}\big)]R'+n(n-1)e^{-2b}\nonumber\\
&&[\{(2a''b'+a'^2b'-a'b'^2+\frac{4a'b'}{r}+\frac{8a'}{r^2}-\frac{4b'^2}{r}-\frac{4a'e^b}{r^2}-\frac{4b'e^b}{r^2} -\frac{8e^b}{r^3}+\frac{2a'}{r^2}+\frac{4b'}{r^2}+\frac{8}{r^3})-\nonumber\\
&&r(2a''b'+a'^2b'-a'b'^2-\frac{4b'^2}{r})\}-\omega\big(a''a'+a''b'-a''-a'^2+a'b'+\frac{a'^3}{2}-\frac{4a'}{r}-\nonumber\\
&&\frac{4b'e^b}{r^2}+\frac{8a''}{r}+\frac{6a'^2}{r}-\frac{13b'}{r^2}-\frac{2b'^2}{r}-\frac{17a'}{r^2}-\frac{18e^b}{r^3}+\frac{18}{r^3}\big)]G^{n-2}G'-\frac{e^{-b}}{r^2}\nonumber\\
&&[2\lambda(\frac{a''r^2}{2}+\frac{a'^2r^2}{4}-\frac{a'b'r^2}{4}-b'r)+\omega(\frac{a''r^2}{2}+\frac{a'^2r^2}{2}-\frac{a'b'r^2}{4}+a'r)](1+2\lambda R)-\nonumber\\
&&\frac{ne^{-2b}}{r^2}(1+\omega)\{(1-e^b)(a'^2+2a''-a'b')- 2a'b'\}G^{n-1}+(1+\omega)\frac{1}{2}(R+\lambda R^2+G^n)=0.\nonumber\\
\end{eqnarray}
\begin{figure}\center
\begin{tabular}{cccc}
\epsfig{file=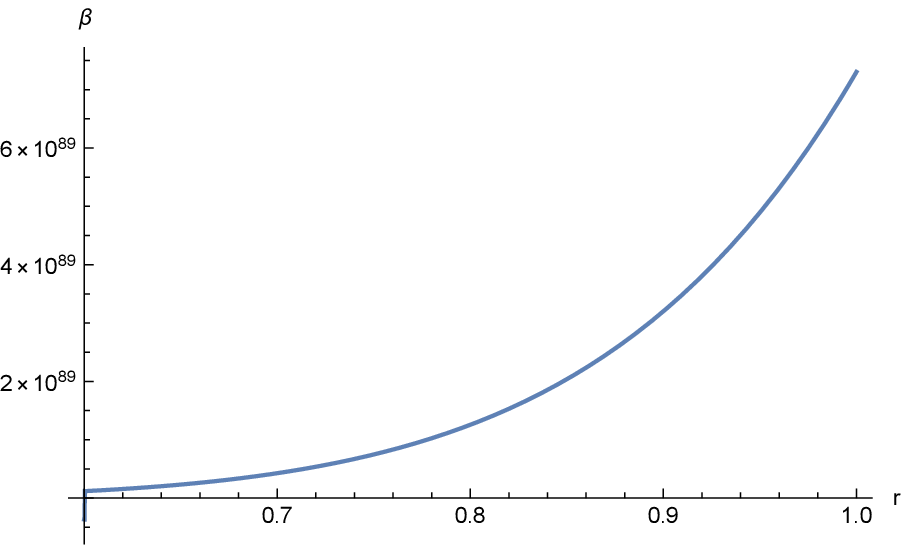,width=0.35\linewidth} &
\epsfig{file=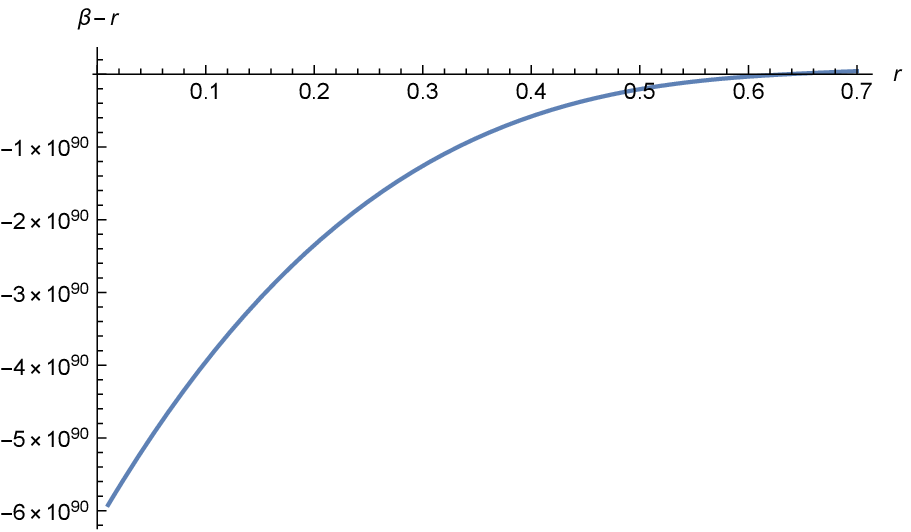,width=0.35\linewidth} &\\
\epsfig{file=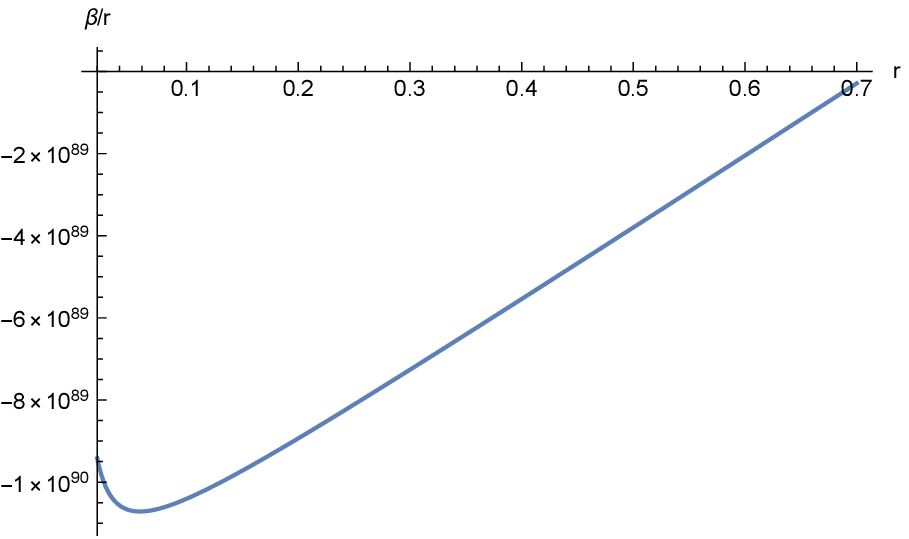,width=0.35\linewidth} &
\epsfig{file=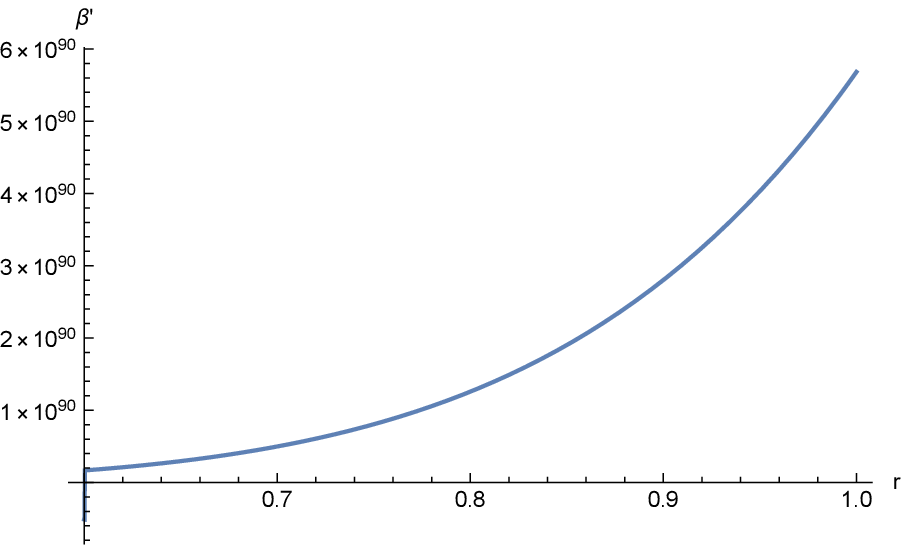,width=0.35\linewidth}& \\
\end{tabular}
\caption{For barotropic fliud, behavior of $\beta(r)$, $\beta(r)-r$, $\beta(r)/r$ and $\beta'(r)$  with radial coordinate $r~~(Km)$. Here, we set  $n=2$, $\eta=5$, $\omega=0.01$ and $\lambda=1$}\center
\end{figure}
Again we convert the equation (\ref{10}) in terms of $\beta(r)$ and get a complicated and highly nonlinear differential equation which is difficult to solve analytically. Therefore, here we also need to apply some suitable numerical techniques to investigate the geometry of $\beta(r)$. The information thus obtained is depicted in  Fig. $5$ which shows $\beta(r)$  as an increasing function. Furthermore, the behavior of $\beta(r)-r$, $\beta(r)/r$ and $\beta'(r)$  with radial coordinate $r$ has been shown in Fig. $5$ which is similar as in isotropic case. The throat is located at $r_0=0.716455$ as the curve $\beta(r)-r \rightarrow 0$. It can be observed from the graph of $\beta'(r)$ that $\beta'(r_0)<1$.
\begin{figure}\center
\begin{tabular}{cccc}
\epsfig{file=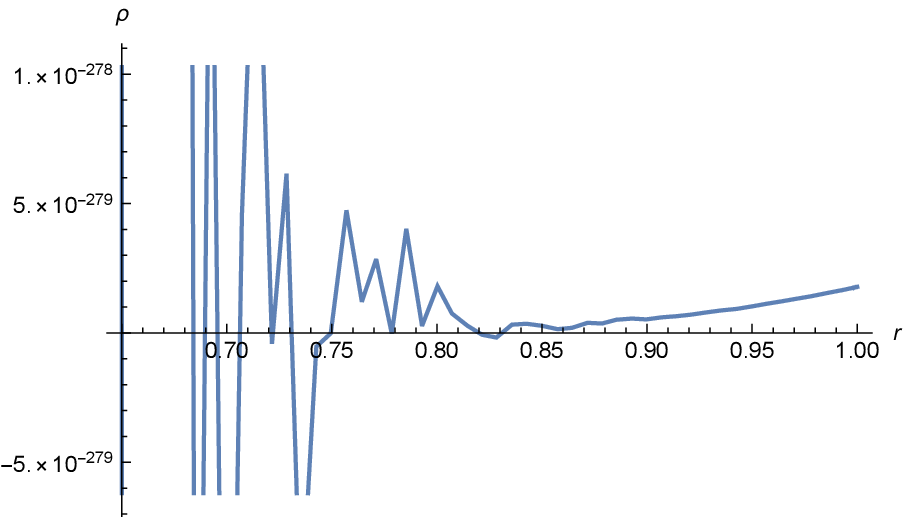,width=0.35\linewidth} &
\epsfig{file=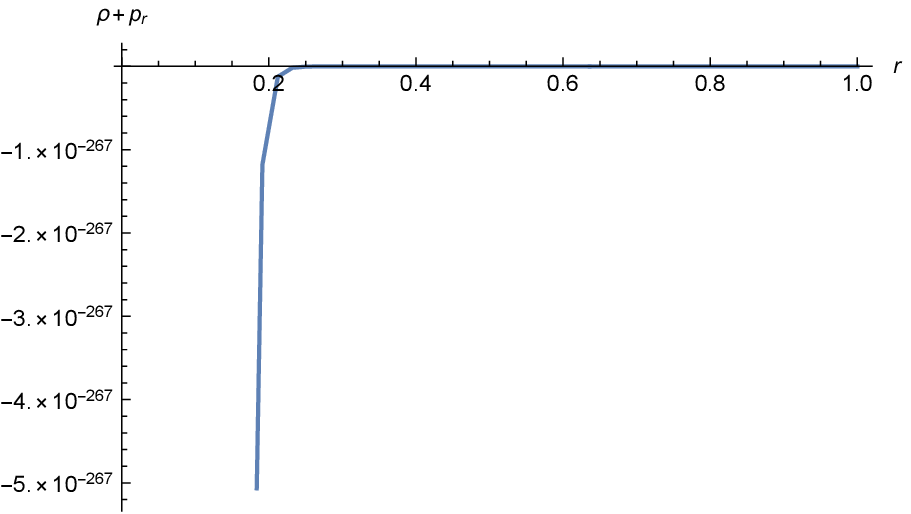,width=0.35\linewidth} &
\epsfig{file=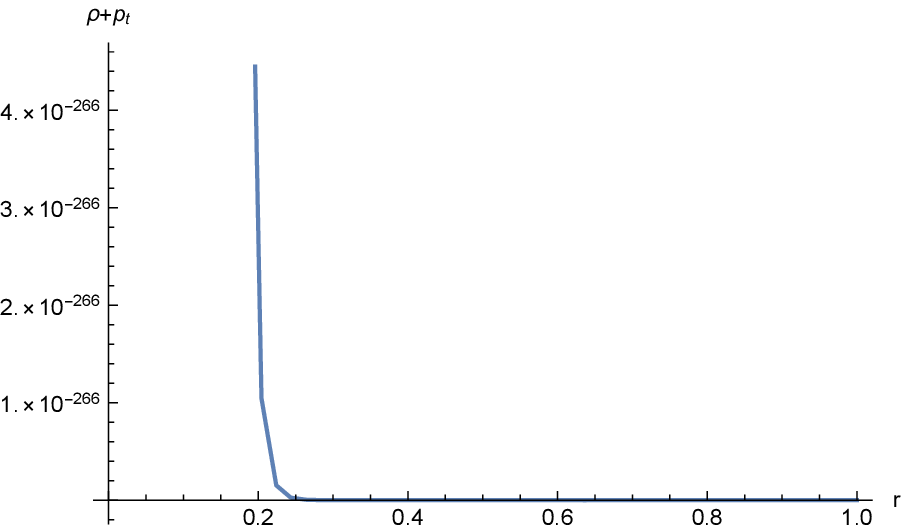,width=0.35\linewidth} \\
\end{tabular}
\caption{Behaviour of $\rho~~(MeV/fm^3)$, $\rho+p_r~~(MeV/fm^3)$ and $\rho+p_t~~(MeV/fm^3)$ with radial coordinate $r~~(Km)$ for barotropic fluid}\center
\end{figure}
The qualitative evolution of $\rho$, $\rho+p_r$ and $\rho + p_t$ has been shown in Fig. $6$. We observe the validity of both WEC and NEC in small region when $0.65\leq r \leq1$. So a realistic wormhole is possible here.

\section{Concluding Remarks}

In this present manuscript, we have investigated some stimulating and useful results about the existence of wormhole geometries through the ordinary matter distributions in  $f(R,G)$ gravity, by considering its particular model $f(R,G)= R+\lambda R^2+G^n$. For this purpose, we have considered anisotropic, barotropic and isotropic fluids with more emphasis on anisotropic fluid as this is the more general case than the other two fluids. We have examined the behavior of the energy conditions, WEC and NEC, done with graphical analysis. In literature, different techniques are being used for discussion of wormhole solutions. We have considered some specific forms of the shape function $\beta$ and the non-constant red-shift function $a$ in the field equations. Further, we have explored the expressions for important components $\rho+ p_r$ and $\rho + p_t$ of the energy conditions, which play dynamic and decisive role in establishing any opinion about possible presence of feasible regions for the wormhole solutions in modified $f(R,G)$ gravity.

Working with anisotropic case in modified $f(R,G)$ gravity, we obtain quite complicated and highly non-linear field equations involving six unknown quantities. Therefore, we face almost an impossible situation to have explicit form for the shape function from the field equations. We are left with the other option of adopting some forms of these two unknowns (shape function and red-shift function) under some assumptions according to the situation. By assigning different values to the parameters occurring in the expressions for the energy bounds, we have concluded the possible existence of wormhole solutions through comprehensive graphical analysis. It has been observed throughout these graphical analysis, WEC and NEC are widely violated for the anisotropic matter content. We have encountered with large feasible regions both for WEC and NEC independently and we also found out the feasible regions common to both WEC and NEC to support the wormhole solutions threaded by the non-exotic matter for different selections of the parametric values $m,~n$ and $\lambda$. We observe some tiny feasible areas for the existence of wormhole solutions close to the wormhole throat. However, while investigating the other two cases of isotropic and barotropic matter contents, the situation is a little bit less complicated as compare to the previous case as far as is the shape function concerned. For both these cases, numerical solutions of the differential equation are found using some suitable numerical technique. With the help of graphical behavior of the shape function, all the essential requirements like asymptotically flat and flaring out conditions are satisfied which confirms the fact that the obtained wormhole is viable. It is stated that the energy conditions are also violated for the isotropic and barotropic cases, while wormhole solutions nearby the throat of the wormhole are found.
Conclusively, the analysis confirms that the wormhole solutions do exist in $f(R,G)$ theory under some reasonable circumstances and proposed $f(R,G)$ gravity model. It may be an interesting work to explore wormhole solutions by considering some other bivariate models in $f(R,G)$ gravity. It is important to  mention here that the results of this study match with the findings in (Sharif and Fatima 2015) for  $f_1(R)=0$ and $n=2$.

\end{document}